\documentclass[pra,onecolumn,superscriptaddress,aps]{revtex4}
\usepackage{amsfonts}
\usepackage{amssymb,latexsym,amsmath}
\usepackage{times}
\usepackage[caption=false]{subfig}
\usepackage[usenames]{color}
\usepackage{graphicx,tikz}
\usepackage{fancyvrb}

\begin{document}

\title{Rogue Waves as Self-Similar Solutions on a Background: A Direct Calculation}

\author{C. B. Ward}
\affiliation{Department of Mathematics and Statistics, University of
	Massachusetts, Amherst MA 01003-4515, USA}

\author{P. G. Kevrekidis}
\affiliation{Department of Mathematics and Statistics, University of
	Massachusetts, Amherst MA 01003-4515, USA}


\begin{abstract}
  In the present work, we explore the possibility of developing
  rogue waves as exact solutions of some nonlinear
  dispersive equations, such as  the nonlinear Schr{\"o}dinger
  equation, but also, in a similar vein,
  the Hirota, Davey-Stewartson, and Zakharov
  models. The solutions that we find are ones previously
  identified through different methods. Nevertheless, they
  highlight an important aspect of these structures, namely
  their self-similarity. They thus offer an {alternative}
  tool in the very sparse (outside of the inverse scattering
  method) toolbox of attempting to identify {analytically} (or computationally)
  rogue wave solutions. This methodology is importantly independent
  of the notion of integrability. An additional nontrivial
  motivation for such a formulation is that it offers a
  frame in which the rogue waves are stationary. It is conceivable
  that in this frame one could perform a proper stability
  analysis of the structures.
\end{abstract}

\maketitle
        
\section{Introduction: Motivation \& Approach}

The study of waves that are characterized by extreme
amplitudes, and which are often referred to as
freak or rogue structures has gained considerable
traction over the last decade. This can, arguably,
to a large degree be attributed to the exploration
of the relevant waveforms in a variety of experimental
realizations spanning diverse areas of physics,
starting from the traditional field of
hydrodynamics~\cite{hydro,hydro2,hydro3}, but
also extending to numerous other areas. These
include, but are not limited to, nonlinear
optics~\cite{opt1,opt2,opt3,opt4,opt5,laser}, superfluid helium~\cite{He},
as well as plasmas~\cite{plasma}. These multifaceted
experimental studies have, in turn, triggered a wide
range of theoretical explorations which by now have
been summarized in a series of reviews~\cite{onorato,solli2,yan_rev},
but also importantly in a series of books on this
research theme~\cite{k2a,k2b,k2c,k2d}.

The nonlinear Schr{\"o}dinger equation (NLS)~\cite{sulem,ablowitz}
has, arguably, been a principal vehicle of choice for the
exploration of rogue waves, with the so-called
Peregrine soliton~\cite{H_Peregrine}, the Akhmediev breather (AB)~\cite{akh},
and the Kuznetsov-Ma~\cite{kuz,ma} (KM) soliton playing a crucial
role in the corresponding studies as the prototypical structures
of relevance; in this context, the work of Dysthe-Trulsen
is also worth mentioning~\cite{dt}. Nevertheless, these
waveforms have emerged from toolboxes associated with
integrability theory and inverse scattering. This naturally
prompts the question of accessibility of relevant
solutions {\it beyond} the integrable limit and of
their persistence under non-integrable perturbations which
are especially important in numerous, among the above
mentioned, applications. In this vein, analytical tools
based on perturbative (exact or approximate) solutions~\cite{Ank1},
as well as on persistence criteria for such
waveforms~\cite{calinibook,pla} have been proposed.
In a recent study~\cite{arxiv},
we developed a complementary approach to the above
analytical ones attempting to identify rogue waves
via a numerically developed toolbox which traces the
existence of such structures as a steady-state problem
in space-time (i.e., considering time as a space-like variable and
performing a conjugate-gradient minimization to obtain the wave
pattern). Our methodology was successful in establishing the
persistence of Peregrine-like patterns, which are localized
in both space and time, in generalized NLS models, for instance
featuring a different than cubic nonlinearity.

Here, we would like to take a different path in the vein
of offering a different kind of theoretical
(but possibly also computational) tool
towards the identification of rogue wave structures.
We are motivated by two facts: on the one hand, the
availability of analytical techniques for finding
rogue waves is largely limited to methods stemming
from inverse scattering transform (IST) and integrable
systems. Nevertheless, in many of the realistic non-integrable
systems where such extreme waves appear the IST formulation
is not available or applicable. Moreover, from a structural perspective,
rogue waves are self-similar in their functional form,
even though, contrary to what is the case for other such
solutions~\cite{sulem}, they do not blow up in finite time.
In light of these developments, we offer an unprecedented
approach on the basis of self-similarity to capture
such solutions. We should caution the reader from the get
go that our aim at least in the present work is not to
identify solutions that have not appeared before
in the literature. All the solutions that we will present
have been previously found via methods related, in
some way or other, to integrability. Rather, we offer
a different perspective of these solutions that is
{\it not} bound by integrability and can, in principle,
be generalized to models that possess freak waves, yet
are not known or expected to be integrable. Moreover, as
we will argue in our Future Challenges section, this
aspect prompts the potential consideration of the stability
of such solutions in the self-similar frame. 

\section{Self-Similar Calculation for the NLS case}

Let us start with the prototypical NLS model of the form:
\begin{eqnarray}
  i u_t =-\frac{1}{2} u_{xx} - |u|^2 u + u,
  \label{sseq1}
\end{eqnarray}
whereby we have already set the background at unity without
loss of generality. Now let us recall that the Peregrine
waveform emerges against the backdrop of a constant background
$u=1+ w$. Then, the equation for the new variable $w(x,t)$
reads:
\begin{eqnarray}
  i w_t=-\frac{1}{2} w_{xx} - |w|^2 w - (w+w^{\star}) -
  (2 |w|^2 + w^2).
  \label{sseq2}
\end{eqnarray}
The star here denotes complex conjugation.
We now follow the so-called MN-dynamics self-similar framework
which has been detailed in a number of different
publications~\cite{mnv1,mn_nls,mn_csinf} (and is also tantamount
to methods generally applied for obtaining self-similar solutions;
see, e.g.,~\cite{baren}). More specifically, we use $w(x,t)=A(t)
F(\frac{x}{L(t)},t)$. This leads to the equation:
\begin{eqnarray}
  &i& \left(A_t F + A F_t - A \xi F_{\xi} \frac{L_t}{L} \right) =
  -\frac{1}{2} \frac{A}{L^2} F_{\xi  \xi} - |A|^2 A |F|^2 F
  \nonumber
  \\
  &-&
  (A F + A^{\star} F^{\star}) - (2 |A|^2 |F|^2 + A^2 F^2).
  \label{sseq3}
\end{eqnarray}
In the above equation, $\xi=x/L(t)$.
We will now seek self-similar solutions, by making two
important assumptions. First, we will assume that the
solutions are {\it stationary in the self-similar frame.}
This is not a restricting assumption: it simply demands that
we have self-similar solutions as such. The second assumption
is that we will assume $F$ to be a real profile. A priori,
this is not mandated, nevertheless an understanding of
the inner workings of the method suggests that it would not
be possible to scale terms like the ones of the second
line of Eq.~(\ref{sseq3}) barring such a restriction.
Effectively, we assign all the complex phase dependence
in $A(t)$. Using on the basis of the above $F_t=0$ and
also decomposing $A(t)=R e^{i \theta}$, we obtain the following
terms:
\begin{eqnarray}
  &i& \left( \frac{R_t}{R} F- \xi F_{\xi} \frac{L_t}{L} \right)
  -\theta_t F = -\frac{1}{2 L^2} F_{\xi  \xi} - R^2 F^3
  \nonumber
  \\
  &-& (1 + e^{-2 i \theta}) F - F^2 R (2 e^{-i \theta}  + e^{i \theta}).
  \label{sseq4}
\end{eqnarray}
Expressing the last two parentheses as a function of $R$ and
$\theta$ (and $F$) and splitting real and imaginary parts, we inherit
the real part equation:
\begin{eqnarray}
  -\theta_t F=-\frac{1}{2 L^2} F_{\xi  \xi} - R^2 F^3 -2 \cos^2(\theta) F
  -3 R \cos(\theta) F^2.
  \label{sseq5}
\end{eqnarray}
For self-similarity to work here in the corresponding self-similar frame,
we need the
time dependent terms to cancel and independently the time-independent
terms to do the same. This leads to the choice
\begin{eqnarray}
  R^2=\frac{1}{L^2}=2 R \cos(\theta), \quad 2 \cos^2(\theta)=\theta_t.
  \label{sseq6}
\end{eqnarray}
The latter, in turn, yields $\tan(\theta)=2 (C + t)$, where $C$ shifting
the origin of time can be set to $0$ without loss of generality.
Then $\cos^2(\theta)=1/(1+4 t^2)$ leading to
\begin{eqnarray}
  L=\frac{1}{R}=\frac{\sqrt{1+ 4 t^2}}{2} \Rightarrow
  A= \frac{2}{1-2 i t}.
  \label{sseq7}
\end{eqnarray}
Having found the temporal prefactor, the resulting real equation
for the wave profile now reads:
\begin{eqnarray}
  -\frac{1}{2} F_{\xi \xi}- F^3 - \frac{3}{2} F^2=0.
  \label{sseq8}
\end{eqnarray}
Using the above Eqs.~(\ref{sseq6})-(\ref{sseq7}) to simplify
the imaginary part of the Eq.~(\ref{sseq4}) canceling
from all the terms a factor of $\sin(2 \theta)$, we retrieve
a first-order equation for the same profile, namely:
\begin{eqnarray}
  \xi F_{\xi}= -2 F - F^2.
  \label{sseq9}
\end{eqnarray}
We can solve this first order ODE to obtain the solution
\begin{eqnarray}
  F(\xi)=- \frac{2}{1+ D \xi^2};
  \label{sseq10}
\end{eqnarray}
now a direct substitution of this waveform in Eq.~(\ref{sseq8})
yields straightforwardly that the solution is valid only for $D=1$, hence
we obtain $F=-2/(1+\xi^2)$. We now reconstruct our solution:
\begin{eqnarray}
  u=1 + W= 1 + \frac{2}{1-2 i t} \left(-\frac{2}{1+\xi^2}\right)
  =1 - 4 \frac{1 + 2 i t}{1+ 4 x^2 + 4 t^2},
  \label{sseq11}
\end{eqnarray}
which naturally retrieves the well-known Peregrine structure~\cite{H_Peregrine}.
It is important to make some remarks here:
\begin{itemize}
\item The expressions of Eq.~(\ref{sseq11}) bring forth the self-similar
  nature of the Peregrine, which, in our view, seems to have been
  overlooked in the literature. Factoring out the time dependence as
  a complex factor ($\propto 1/(1-2 i t)$), one is left with an
  effective Lorentzian self-similar waveform which is at the
  heart of the self-similarity-based
  calculation of the structure as a steady state
  solution of the relevant formulation.
\item Nevertheless, 
  there are some nontrivial differences of this calculation
  from other similar calculations, e.g., in the above mentioned
  references such as~\cite{mnv1,mn_nls,mn_csinf}. Here, when
  separating, for instance, the real part of the solution, there
  are both time-dependent and time-independent terms
  and these need to be balanced out between them separately
  (the former on their own, and the latter on their own).
  Perhaps even more importantly, the real and imaginary parts
  yield 2 distinct differential equations that have a particular
  solution in common that needs to be singled out upon a suitable
  selection of a compatibility constant.
\end{itemize}
Nevertheless, the above procedure can be utilized whenever
it may be believed (e.g., motivated from numerical
or physical experiments) that extreme waves may exist
in a certain system. The methodology only hinges on identifying
such a self-similar solution on top of a background and in
no way utilizes the integrable structure of the model. In a
sense, it is an analogous calculation (for rogue waves)
to the reduction of the NLS, when looking for its standing
waves, to a Duffing oscillator whose homoclinic or heteroclinic
connections correspond to the bright or dark solitons
respectively. To the best of our knowledge such as
an ODE-reduction-based calculation has not previously appeared
in the context of rogue waves. To illustrate that this
viewpoint for the consideration of rogue wave patterns can be
used in other (admittedly related) examples 
beyond ``just NLS'', we consider now a series of generalizing cases,
including the Hirota model in 1+1 dimensions and the Davey-Stewartson
and Zakharov models in higher (i.e., 2+1) dimensional settings. 
The Hirota model incorporates effects such as the third order
dispersion and the time-delay correction to the cubic nonlinearity,
as discussed, e.g., in~\cite{sedletskii}.
The Davey-Stewartson model is a relevant one for the examination
of the evolution of a wave-packet in a (2+1)-dimensional setting
for water of finite depth~\cite{ablowitz1}; here, we consider
the setting of large surface tension in the form of the so-called
DSI model.
Finally, the Zakharov equation was derived in~\cite{zakharov} as a prototypical
integrable model in (2+1)-dimensions. 

\section{Going Beyond NLS: Other Models}

\subsection{Hirota Equation}
The Hirota equation:
\begin{equation}
i u_t + \frac{1}{2} u_{xx} + |u|^2 u - \alpha i u_{xxx} -6\alpha i |u|^2 u_x =0,
\end{equation}
where $\alpha$ is an arbitrary constant, has a rogue wave solution
of the form~\citep{Hirota}
\begin{equation}
u(x,t) = e^{it} \bigg( 1-\frac{4(1+2it)}{1+4(x+6\alpha t)^2+4t^2}\bigg)
\end{equation}
Again, by factoring out $1+4t^2$ from the bottom, we may write
\begin{equation}
u(x,t) = e^{it} \bigg( 1+\frac{2(1+2it)}{1+4t^2} \cdot \frac{-2}{1+(\frac{2(x+6 \alpha t)}{\sqrt{1+4t^2}})^2}\bigg).
\end{equation}
where
\[
\xi =\frac{2(x+6 \alpha t)}{\sqrt{1+4t^2}}.
\]
We then see that the rogue wave of the Hirota equation is, in a very natural way, a self-similar, Peregrine-like structure in
a co-traveling reference frame where the coherent structure travels
with speed $6 \alpha$. Hence, it can be retrieved by a similar
calculation as the one above.

\subsection{Davey-Stewartson I}
The $(2+1)$-dimensional Davey-Stewartson I (DSI) equation 

\begin{align}
iu_t &= u_{xx} + u_{yy} + |u|^2 u - 2Qu \\
Q_{xx} - Q_{yy} &= (|u|^2)_{xx},
\end{align}

has a rogue wave solution~\citep{DavyStewartson} of the form:

\begin{align}
u(x,y,t) &=\sqrt{2}\bigg(1-\frac{4(1-2i \omega t)}{1+(k_1x+k_2y)^2+4 \omega^2 t^2} \bigg) \\
Q(x,y,t)&= 1-4k_1^2 \frac{1-(k_1x+k_2y)^2+4 \omega^2 t^2}{(1+(k_1x+k_2y)^2+4 \omega^2 t^2)^2},
\end{align}
where $k_1=p-\frac{1}{p}$, $k_2=p+\frac{1}{p}$, $\omega=p^2+\frac{1}{p^2}$, and $p$ is an arbitrary constant. We note here that this is a line rogue wave, resembling the Peregrine structure extended along a line in the $xy$-plane. We can write this in the self-similar form

\begin{align}
u(x,y,t) &=\sqrt{2}\bigg(1+ \frac{2(1-2i\omega t)}{1+4\omega^2 t^2} \cdot \frac{-2}{1+\xi^2} \bigg) \\
Q(x,y,t)&= 1+\frac{4k_1^2 }{1+4 \omega^2 t^2} \cdot \frac{\xi^2-1}{(\xi^2+1)^2}
\end{align}
where 
\[
\xi = \frac{k_1x+k_2y}{\sqrt{1+4 \omega^2 t^2}}.
\]
Hence, in this case too, the structure can be thought of as being
stationary in a suitable self-similar frame of reference.
\subsection{Zakharov Equation}
The (2+1)-dimensional Zakharov equation assumes the form~\cite{zakharov}:
\begin{eqnarray}
iu_t &= u_{xy} + Qu \\
Q_{y} &= 2(|u|^2)_{x}.
\label{zakh2}
\end{eqnarray}

This model also admits line-type rogue waves~\citep{Zakharov2}, which we will simply give in self-similar form:

\begin{align}
u(x,y,t) &=1+ \frac{2(1+4it)}{1+16 t^2} \cdot \frac{-2}{1+\xi^2} \\
Q(x,y,t)&= \frac{16}{1+16t^2} \cdot \frac{\xi^2-1}{(\xi^2+1)^2}
\end{align}
where 
\[
\xi = \frac{2(x-y)}{\sqrt{1+16 t^2}}.
\]
Lastly, we remark that a change in the time scale will change the $\sqrt{1+16t^2}$ to $\sqrt{1+4t^2}$, as in the other examples. This is due to a rescaling
of time imposed effectively by the prefactor within Eq.~(\ref{zakh2}).

\section{Conclusions \& Future Work}

In the present work we have revisited the examination of rogue
wave structures in the context of dispersive nonlinear models.
We have argued that while the IST and related techniques (including
e.g. the Darboux transformation etc.) provide valuable tools for
identifying such solutions in the realm of integrable models, this
methodology is limited in comparison to more realistic models that bear
non-integrable perturbations and for which experimental or numerical
observations suggest that the structures may persist. A perturbative
framework either analytically~\cite{Ank1,pla,calinibook} or
numerically~\cite{arxiv} may be useful in such scenarios. Nevertheless,
we argue that a potentially valuable complementary perspective
is that of looking at rogue wave patterns as self-similar solutions
that are associated with a (potentially complex) time-dependent
prefactor and a self-similar (e.g. in the NLS case, Lorentzian)
profile, arising against the backdrop of a constant, non-vanishing
background. Seeking these solutions through a self-similar
type of methodology, as was done for some prototypical case
examples herein, enables a way to tackle such solutions that is not
bound by the limitations of integrable models and can instead
be applied to a wider range of systems.

That being said, in the present proof of principle exposition we
have only retrieved case examples where the existence of such
waveforms was already identified by integrable structure means,
in order to illustrate the ability of the method to capture such
waveforms. However, it would be of particular interest to attempt
a similar search in model examples where rogue structures are
expected to exist, yet the absence of integrability does not
allow for their identification. This is one of the key challenges
of the method towards future work. A related challenge lies
in the potential for consideration of stability features
in the self-similar frame. This was done, e.g., in~\cite{mn_nls},
but also, importantly, in a series of works of~\cite{wit1,wit2}; see
also~\cite{ray}.
These efforts have not only identified the spectra of self-similar
waveforms (which are steady in the self-similar frame); they have
importantly made a substantial effort to ``reinterpret'' the
spectra of the latter setting into the original frame. An
important example of this class is that certain symmetries
(such as, e.g., the potential shift of the collapse time
in non-autonomous systems) may amount to eigendirections
appearing as unstable, which, yet, are not so due to the
existence of the corresponding symmetry (in the original frame).
This is yet to be done for the Peregrine soliton of the NLS
and related waveforms. We have attempted this and have been
hindered by technical complications having to do with the
nature of the emerging terms in Eq.~(\ref{sseq3}). This is
the same complication that we encountered previously in that
some of the terms are autonomous and some are not. The potential
existence of a systematic way to bypass this complication would
pave a systematic way for identifying (and subsequently
reinterpreting in the spirit of~\cite{wit1,wit2}) the spectra
of extreme wave events, a feature crucial for formulating
a more precise notion of their stability. Up to now the
latter has been explored in either a somewhat empirical
(and often mathematically not suitably substantiated) way
or in the form of a limiting procedure of, e.g.,
periodic states; see the relevant discussion of~\cite{kmpaper}.
Nevertheless, such a direct approach as proposed here would
be fundamentally superior to the current state of the art,
in our view, and hence constitutes a particularly worthwhile topic
for future study.


\begin{thebibliography}{99}

\bibitem{hydro} A. Chabchoub, N. P. Hoffmann, and N. Akhmediev,
Phys. Rev. Lett. {\bf 106}, 204502 (2011).

\bibitem{hydro2} A. Chabchoub, N. Hoffmann, M. Onorato,
and N. Akhmediev, Phys. Rev. X {\bf 2}, 011015 (2012).

\bibitem{hydro3} A. Chabchoub and M. Fink, Phys. Rev. Lett. {\bf 112}, 124101 (2014).

  \bibitem{opt1} D. R. Solli, C. Ropers, P. Koonath, and B. Jalali,
Nature {\bf 450}, 1054 (2007).

\bibitem{opt2} B. Kibler {\it et al.}, Nature Phys. {\bf 6}, 790 (2010).

\bibitem{opt3} B. Kibler {\it et al.}, Sci. Rep. {\bf 2}, 463 (2012).

\bibitem{opt4} J. M. Dudley, F. Dias, M. Erkintalo, and G. Genty,
Nat. Photon. {\bf 8}, 755
(2014).

\bibitem{opt5} B. Frisquet {\it et al.}, Sci. Rep. {\bf 6}, 20785 (2016).

\bibitem{laser} C. Lecaplain, Ph. Grelu, J. M. Soto-Crespo, and N. Akhmediev,
  Phys. Rev. Lett. 108, 233901 (2012).

\bibitem{He} A. N. Ganshin, V. B. Efimov, G. V. Kolmakov, L. P. Mezhov-Deglin,
and P. V. E. McClintock, Phys. Rev. Lett. {\bf 101}, 065303 (2008).

 \bibitem{plasma} H. Bailung, S. K. Sharma, and Y. Nakamura,
Phys. Rev. Lett. {\bf 107}, 255005 (2011).

 
\bibitem{onorato} M. Onorato, S. Residori, U. Bortolozzo, A. Montinad, and  F. T. Arecchi,
Phys. Rep. {\bf 528}, 47 (2013).

\bibitem{solli2} P. T. S. DeVore, D. R. Solli, D. Borlaug, C. Ropers, and B. Jalali,
J. Opt. {\bf 15}, 0640031 (2013).

  \bibitem{yan_rev} Z. Yan,
J. Phys. Conf. Ser. {\bf 400}, 012084 (2012).

 \bibitem{k2a} E. Pelinovsky and C. Kharif (eds.),
{\it Extreme Ocean Waves} (Springer, NY, 2008).

\bibitem{k2b} C. Kharif, E. Pelinovsky, and A. Slunyaev,
{\it Rogue Waves in the Ocean} (Springer, NY, 2009).

\bibitem{k2c} A. R. Osborne, {\it Nonlinear Ocean Waves and the Inverse Scattering
Transform} (Academic Press, Amsterdam, 2010).

\bibitem{k2d} M. Onorato, S. Residori, and F. Baronio,
  {\it Rogue and Shock Waves in Nonlinear Dispersive Media},
  Springer-Verlag (Heidelberg, 2016).

\bibitem{sulem} C. Sulem and P.L. Sulem, 
\newblock  {\it The Nonlinear Schr{\"o}dinger Equation}, 
Springer-Verlag (New York, 1999).


\bibitem{ablowitz} M.J. Ablowitz, B. Prinari and A.D. Trubatch,
{\it Discrete and Continuous Nonlinear Schr{\"o}dinger Systems},
Cambridge University Press (Cambridge, 2004).

  
  

  \bibitem{H_Peregrine} D.~H.~Peregrine, %
J. Austral. Math. Soc. B \textbf{25}, \ 16 \ (1983).

\bibitem{akh} N. N. Akhmediev, V. M. Eleonskii, and N. E. Kulagin,
Theor. Math. Phys. {\bf 72}, 809 (1987).


\bibitem{kuz} E. A. Kuznetsov, Sov. Phys.-Dokl. {\bf 22}, 507 (1977).

\bibitem{ma} Ya. C. Ma, Stud. Appl. Math. {\bf 60}, 43 (1979).


\bibitem{dt} K. B. Dysthe and K. Trulsen, Phys. Scr. {\bf T82}, 48 (1999).


  
  \bibitem{Ank1} A Ankiewicz, N Devine, N Ahkmediev,
Physics Letters A {\bf 373} (2009) 3997

  
  \bibitem{calinibook} A. Calini and C. M. Schober,
pp.~31--51 in Ref.~\cite{k2a}.

\bibitem{pla} A. Calini, C.M. Schober,
  Phys. Lett. A {\bf 298}, 335 (2002).

\bibitem{arxiv} C. B. Ward, P. G. Kevrekidis, N. Whitaker,
  arXiv:1712.03292.


\bibitem{mnv1} D.G. Aronson, S.I. Betelu, I. G. Kevrekidis,
  arXiv:nlin/0111055.

\bibitem{mn_nls} C.I. Siettos, I.G. Kevrekidis, P.G. Kevrekidis,
  Nonlinearity {\bf 16}, 497 (2003).

\bibitem{mn_csinf} P.G. Kevrekidis, C.I. Siettos, Y.G. Kevrekidis, Nature
  Comms. {\bf 8}, 1562 (2017).

\bibitem{baren} G.I. Barenblatt,
  {\it Scaling, self-similarity and intermediate asymptotics},
  Cambridge University Press (Cambridge, 1996).

\bibitem{sedletskii} Yu.V. Sedletsky,
  J. Exp. Theor. Phys. {\bf 97}, 180 (2003)

  
\bibitem{ablowitz1} M. J. Ablowitz and H. Segur,
  {\it Solitons and the Inverse Scattering Transform}
  (SIAM, Philadelphia, 1981).
  

\bibitem{zakharov}
  V. E. Zakharov, Solitons
  in Topics Curr. Phys., Vol. 17, R. K. Bullough and P. J. Caudrey (Eds.)
  Springer-Verlag (Berlin 1980).
  
 \bibitem{Hirota} A. Ankiewicz, J. M. Soto-Crespo, and N. Akhmediev, Physical Review E {\bf 81}, 046602 (2010).

   \bibitem{DavyStewartson} Y. Ohta and J. K. Yang,  Physical Review E  {\bf 86}, 036604 (2012).

\bibitem{Zakharov2} J. Rao, L. Wang, W. Liu, and J. He,
Theoretical and Mathematical Physics {\bf 193}, 1783 (2017)

     
  
\bibitem{wit1} A.J. Bernoff, T.P. Witelski,
  Appl. Math. Lett. {\bf 15}, 599 (2002).

\bibitem{wit2} A.J. Bernoff, T.P. Witelski,
  J. Eng. Math {\bf 66}, 11 (2010).


\bibitem{ray} S. Ray, R.C. Viesca,
  J. Geophys. Res.: Solid Earth {\bf 122}, 8214 (2017).
  
\bibitem{kmpaper} J. Cuevas-Maraver, P. G. Kevrekidis, D. J. Frantzeskakis, N. I. Karachalios, M. Haragus, and G. James
  Phys. Rev. E 96, 012202 (2017)
  


 
  \end{thebibliography}
\end{document}